\documentclass[aip,jcp,reprint]{revtex4-1}

\usepackage{amsmath,amssymb,graphicx,color}
\usepackage[colorlinks=true,citecolor=red,linkcolor=blue]{hyperref}

\newcommand{\lit}{\mbox{Li$_{2}$\ }}
\newcommand{\litp}{\mbox{Li$_{2}^{+}$\ }}
\newcommand{\wn}{\mbox{cm$^{-1}$}}
\def\mathbi#1{\textbf{\em #1}}

\begin{document}

\title{Extracting spectroscopic molecular parameters from short pulse photo-electron angular distributions}

\author{R.\,Chamakhi}
\affiliation{LSAMA, Department of Physics, Faculty of Sciences of Tunis, University of Tunis-El Manar, 2092 Tunis, Tunisia.}

\author{R.\,Puthumpally-Joseph}
\affiliation{Laboratoire Interdisciplinaire Carnot de Bourgogne, CNRS, Universit\'e Bourgogne Franche-Comt\'e, BP 47870, 21078 Dijon, France.}
\affiliation{Institut des Sciences Mol\'eculaires d'Orsay, CNRS, Univ. Paris-Sud, Universit\'e Paris-Saclay, 91405 Orsay, France.}

\author{M.\,Telmini}
\affiliation{LSAMA, Department of Physics, Faculty of Sciences of Tunis, University of Tunis-El Manar, 2092 Tunis, Tunisia.}

\author{E.\,Charron}
\affiliation{Institut des Sciences Mol\'eculaires d'Orsay, CNRS, Univ. Paris-Sud, Universit\'e Paris-Saclay, 91405 Orsay, France.}

\begin{abstract}
Using a quantum wave packet simulation including the nuclear and electronic degrees of freedom, we investigate the femtosecond and picosecond energy- and angle-resolved photoelectron spectra of the E($^1\Sigma_g^+$) electronic state of {Li$_{2}$}. We find that the angular distributions of the emitted photoelectrons depend strongly on the pulse duration in the regime of ultrashort laser pulses. This effect is illustrated by the extraction of a time-dependent asymmetry parameter whose variation with pulse duration can be explained by an incoherent average over different ion rotational quantum numbers. We then derive for the variation of the asymmetry parameter a simple analytical formula, which can be used to extract the asymptotic CW asymmetry parameters of individual transitions from measurements performed with ultra-short pulses.
\end{abstract}

\maketitle

\section{Introduction}
\label{sec:intro}

The smallest neutral molecules such as alkali dimers for instance have often served as benchmark systems in theoretical or experimental studies of femtosecond molecular dynamics due to their simple electronic structure which can be easily probed with common femtosecond laser systems \cite{JCP.103.7269, PRA.54.R4605, JCP.108.9259, CPL.302.363, APB.71.259, JCP.112.8871, CPL.339.362, JCP.114.1259, JCP.114.10311, PRA.66.043402, CPL.376.457, PRA.68.043409, CPL.402.27, CPL.402.126, CPL.26.073301, CPB.19.033301}. In this context, the lithium dimer has been extensively studied both experimentally \cite{JCP.103.7269, JCP.108.9259, JCP.114.10311, PRA.66.043402, PRA.68.043409, CPL.402.27, CPL.402.126} and theoretically \cite{JCP.114.1259, CPL.26.073301, CPB.19.033301, PRA.74.033407}, and time-resolved photoelectron energy distributions were sometimes used as an efficient probe of the molecular dynamics. These studies are usually based on a three-step photoionization scheme \cite{JCP.114.10311, JCP.103.7269, JCP.108.9259, PRA.66.043402, PRA.68.043409, CPL.402.27, CPL.402.126} where a combination of linearly polarized laser pulses excite, in a two-photon process, either a single transition from the ground electronic state X\,($^1\Sigma_g^+$) of \lit to a pure rovibrational level of the excited E\,($^1\Sigma_g^+$) electronic state, or a series of such transitions. The first excited A\,($^1\Sigma_u^+$) electronic state serves then as an intermediate resonance. The populated rovibrational levels of the E electronic potential are then ionized with an ultrashort laser pulse of linear polarization.

In the present study of \lit photoionization, instead of concentrating our attention on photoelectron energy distributions, we focus on photoelectron angular distributions and we show that a strong modification of these angular distributions happens when the pulse duration varies from the femtosecond to the picosecond regime. We quantify this modification by calculating the asymmetry parameter $\beta$ of the emitted photoelectrons as a function of the pulse duration. Our results are then rationalized using a simple analytical model, and we finally show that this model can be used to extract spectroscopic molecular parameters from measurements of the photoelectron angular distribution performed with ultrashort laser pulses, even if the pulse duration is much shorter than the rotational period of both the neutral molecule and the ion.

The outline of the paper is as follows. In Sec.\,\ref{sec:model} we briefly recall the theoretical model we use for the calculation of the molecular photoionization process using ultrashort laser pulses. Then, in Sec.\,\ref{sec:results}, through results of numerical simulations, we illustrate the specific features of the photoelectron angular distributions and calculate the asymmetry parameter at specific photoelectron energies for different pulse durations. We then show that a simple analytical expression is able to reproduce the numerical results and to predict relevant spectroscopic parameters. The last section finally gives some concluding remarks and perspectives for future work.

\section{Theoretical model}
\label{sec:model}

\begin{figure}[ht!]
\includegraphics[width=8.6cm]{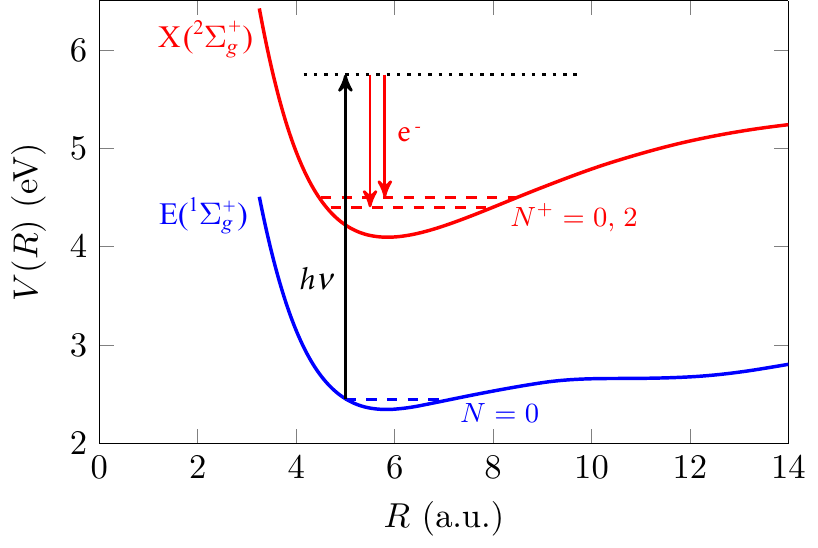}
\caption{(Color online) Potential energy curves involved in the photoionization of \lit as a function of the internuclear distance $R$. The potential energy curve associated with the E$(^1\Sigma_g^+)$ electronic state of \lit is in blue (lower curve), and the potential energy curve associated with the $X(^2\Sigma_g^+)$ of \litp is in red (upper curve). Their associated molecular rotational quantum numbers are denoted by $N$ and $N^+$, respectively.}
\label{fig:PES}
\end{figure}

The present theoretical study is based on a three-step excitation scheme used in several experiments \cite{JCP.103.7269, JCP.108.9259, JCP.114.10311, CPL.402.27, CPL.402.126, PRA.66.043402, PRA.68.043409}, and we concentrate here our attention on the last step which consists in the photoionization of the E$(^1\Sigma_g^+)$ electronic state of  {Li$_{2}$}, as shown in Fig.\,\ref{fig:PES}.

We assume that in the first two steps, a sequence of two CW linearly polarized laser pulses has been used to excite in a two-photon transition a pure rovibrational $(v=0,N=0)$ level on the E$(^1\Sigma_g^+)$ excited electronic potential curve of  {Li$_{2}$}. This kind of technique has already been used with \lit \cite{PRA.66.043402, PRA.68.043409}. Since we assume that the \lit molecule is prepared in its ground rotational level $N=0$, the value of its projection on any quantization axis is $M=0$. This corresponds to an isotropic initial distribution.

We propose to ionize the molecule with the help of a femtosecond or picosecond ionizing pulse. Depending on its exact duration, this laser pulse has a bandwidth $\Delta\omega$ which can encompass the two rotational components $N^+=0$ and $N^+=2$, which can be accessed in the ground electronic state of the ion. On this time scale, describing the ionization process and the induced nuclear dynamics requires a full quantum treatment. Here we recall the particular points of the quantum model that are essential for understanding our forthcoming discussion. For a detailed description of the model, see Refs. [\onlinecite{PRA.74.033407, JCP.138.024108, JCP.144.154109, PRB.95.115406}].

This model was designed to treat the ionization dynamics of neutral alkali dimers, including the induced ro-vibrational dynamics. We follow the electronic dynamics and the nuclear motion by expanding the total molecular wave function as
\begin{eqnarray}
\Psi(\mathbi{r},\mathbi{R},t) & = & 
\psi^{E}(\mathbi{R},t)\,\phi^{E}(r|R)\,Y_{00}(\,\hat{\mathbi{\!r}}\,) + \nonumber\\
& \displaystyle\sum_{m}\int & \psi^{+}_{\ell m}(E,\mathbi{R},t)\,
\phi_{\ell}^{+}(E,r|R)\,Y_{\ell m}(\,\hat{\mathbi{\!r}}\,)\,dE\;\;
\label{Eq:DefWF}
\end{eqnarray}
where $\phi^{E}(r|R)\,Y_{00}(\,\hat{\mathbi{\!r}}\,)$ denotes the electronic wave function associated with the E-state of {Li$_{2}$}. This electronic state of $^1\Sigma_g^+$ symmetry is considered as a 3s$\sigma$ Rydberg state, and its wave function is expressed in the \textit{molecular frame} following Hund's case (b) representation \cite{CUP.2003}. On the other hand, the ionized wave function $\phi_{\ell}^{+}(E,r|R)\,Y_{\ell m}(\,\hat{\mathbi{\!r}}\,)$ is expressed in the \textit{laboratory frame} following Hund's case (d) representation \cite{CUP.2003}. The electron coordinate is denoted by $\mathbi{r}$, and $\mathbi{R}$ denotes the internuclear coordinate. $E$ is the electron asymptotic kinetic energy and $(\ell,m)$ denote the electron angular momentum and its projection in the laboratory frame. Note that the one-photon ionization considered here results in the ejection of a $p$-type electron with $\ell=1$. The sum over $\ell$ is thus omitted in Eq.\,(\ref{Eq:DefWF}). Similar approaches based on such time-dependent expansions have been used in the past for the calculation of time-resolved photoelectron spectra \cite{JCP.110.147, CPL.302.363, JCP.112.8871, JCP.114.1259}.

The photoionization step is performed
by a laser pulse of linear polarization and the polarization axis is chosen as the quantization axis in the laboratory frame. The following selection rule therefore applies for the projections $M$
and $M^+$ of the molecular rotational angular momenta of \lit
and \litp : $M = M^+ + m$. In the present case, we thus have $M^+ + m = 0$. The rotational motion is described by expanding the nuclear wave packets $\psi^{E}(\mathbi{R},t)$ and $\psi^{+}_{\ell m}(E,\mathbi{R},t)$ of Eq.\,(\ref{Eq:DefWF}) as
\begin{subequations}
\label{eq:psiE+}
\begin{eqnarray}
\label{eq:psiE}
\psi^{E}(\mathbi{R},t)
& \!=\! & \psi^{E}_{N,M}(R,t) \, {\cal D}^{N^{\,*}}_{M,0}(\,\hat{\mathbi{\!R}}\,)\\
\label{eq:psi+lm}
\psi^{+}_{\ell m}(E,\mathbi{R},t)
& \!=\! & \!\sum_{N^{+}} \psi_{\ell,m}^{N^+}(E,R,t) \, {\cal D}^{N^{+\,*}}_{M^+,0}(\,\hat{\mathbi{\!R}}\,)
\end{eqnarray}
\end{subequations}
where ${\cal D}^{N^{*}}_{M,\Lambda}(\,\hat{\mathbi{\!R}}\,)$ denote the normalized Wigner rotation matrices \cite{Zare}. Introducing these expansions in the time-dependent Schr\"odinger equation describing the molecule-field interaction and projecting onto the electronic and rotational basis functions yields, in the dipole approximation, a set of coupled differential equations \cite{PRA.74.033407} for the nuclear wave packets $\psi^{E}(\mathbi{R},t)$ and $\psi^{+}_{\ell m}(E,\mathbi{R},t)$ that we solve using the short-time split-operator method\cite{JComputP.47.412}. To integrate the corresponding differential equations we need to define the potential energy curves $V_{E}(R)$ and $V_{+}(R)$ associated with the electronic states of the neutral and of the ion, and the matrix elements which couple the nuclear wave packets evolving on these electronic potential curves. Such matrix elements are given in Ref. [\onlinecite{PRA.74.033407}], and for the potential curves we use the accurate ab-initio values given in Ref. [\onlinecite{CP.92.263}]. Note that the rotational constant of the E state of \lit, $B_E \simeq 0.506\,$\wn, is very close to the one of the ground state of the ion, $B_+ \simeq 0.503\,$\wn. The molecules are subjected to a linearly polarized ionizing field with the classical expression
\begin{equation}
\mathbi{E}(t) = E_0\,f(t)\,\cos(\omega_0 t)\,\hat{\mathbi{\!e}}\,,
\end{equation}
where $E_0$ and $\omega_0$ denote the electric field amplitude and the frequency of the radiation. $\hat{\mathbi{\!e}}$ is the polarization vector. The pulse envelope is finally defined as
\begin{equation}
f(t) = \sin^2(\pi t /(2\tau))\,,
\end{equation}
$\tau$ being the width of the pulse and $2\tau$ its total duration.

The analysis of the photoelectron angular distributions is made by projecting the wave packets $\psi^{+}_{\ell m}(E,\mathbi{R},t_f)$ defined in Eq.\,(\ref{eq:psi+lm}) at the end of the pulse (at time $t_f$) on the energy-normalized solutions of the field-free ionized molecular states represented by the usual expansions on angular momentum states\cite{PRA.49.R641, PRA.51.4824, Messiah}. After integration over the electronic coordinate and over the angular degree of freedom for the nuclei, the angular distribution of the ejected photoelectron at some prescribed asymptotic electron kinetic energy $E=\hbar^2k^2/2m$ and angle $\,\hat{\mathbi{\!k}}\,$ is given by
\begin{equation}
\label{eq:P(E,theta)simple}
P(E,\,\hat{\mathbi{\!k}}\,) = \sum_{N^{+}} \,P_{N^+}(E,\,\hat{\mathbi{\!k}}\,)\,,
\end{equation}
where
\begin{equation}
\label{eq:PNplus(E,theta)simple}
P_{N^+}(E,\,\hat{\mathbi{\!k}}\,) = \sum_{m} \,|Y_{\ell m}(\,\hat{\mathbi{\!k}}\,)|^2
\int |\psi^{N^{+}}_{\ell,m}(E,R,t_f)|^2\,dR\,.
\end{equation}

Integrating $P(E,\,\hat{\mathbi{\!k}}\,)$ over the electron kinetic energy $E$ yields the total angular distribution $P(\,\hat{\mathbi{\!k}}\,)$, while an integration of the same quantity over the angle $\,\hat{\mathbi{\!k}}=(\theta,\phi)$ yields the integrated photoelectron spectrum $P(E)$. In the next section\,\ref{sec:results}, we will discuss both the integrated photoelectron spectra and the angular distributions at specific energies $P(E,\,\hat{\mathbi{\!k}}\,)$ corresponding to different exit channels. In the present case with linear polarization, the angular distributions do not vary with the polar angle $\phi$ and they are just functions of the azimuthal angle $\theta$. We will thus write $P(E,\,\hat{\mathbi{\!k}}\,) = P(E,\theta)$. Compared to our previous study \cite{PRA.74.033407}, which concentrated on kinetic energy distributions only, we put here the emphasis on angular distributions. Note that the theory described here can be generalized to non-isotropic initial conditions by adding a sum over $M$ in Eq.\,(\ref{eq:psiE}). In practice, we use this generalized approach \cite{PRA.74.033407} when $N \neq 0$.

\section{Results}
\label{sec:results}

\subsection{Photoelectron spectra}

Following the scenario shown in Fig.\,\ref{fig:PES}, the second harmonic of a Nd:YAG laser pulse induces photoionization of the molecule at 532\,nm. Depending on the duration of the pulse, between 50\,fs and 15\,ps, the angular distribution may change dramatically because of a strong overlap of the final rotational exit channels. In the following subsections, we will discuss two cases corresponding to the same initial vibrational state $v=0$ in the E-state of the neutral molecule, but with two different initial rotational excitations, $N=0$ and $N=2$.

\subsubsection{Isotropic rotational case $N=0$}

\begin{figure}[ht!]
\includegraphics[width=8.6cm]{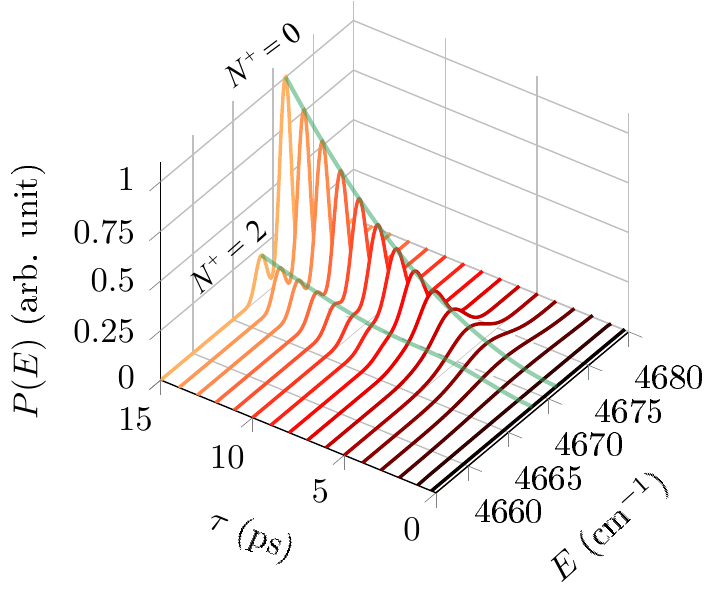}
\caption{(Color online) Photoelectron spectra calculated using a single initial rovibrational level $(v=0,N=0)$ as a function of the electron kinetic energy $E$ in wavenumbers and of the pulse duration $\tau$ in ps. The laser wavelength is $\lambda=532$\,nm (laser Nd:YAG).}
\label{fig:Photoelectron_Spectrum}
\end{figure}

Starting from the initial rovibrational level $v=0$, $N=0$, $M=0$ and taking into account the vibrational selection rule $v_{+}=v$ which corresponds to the highest Frank-Condon factor, there are two different exit channels corresponding to the two accessible ion rotational quantum numbers $N^+=0$ and $N^+=2$. In Fig.\,\ref{fig:Photoelectron_Spectrum}, we display the photoionization probability as a function of the electron kinetic energy for various pulse durations $\tau$ (Full Width at Half Maximum). For very short pulses $(\tau < 5\,\mathrm{ps})$, the spectra show a broad single  peak, while a well-separated  double-peak structure appears for long pulse durations $(\tau > 10\,\mathrm{ps})$. The centers of the peaks correspond to the energies associated with the ion ro-vibrational levels $(v_+=0,N^+=0)$ and $(v_+=0,N^+=2)$, at $E=4671.6$\,\wn  $\,$ and $E=4668.5$\,\wn, respectively.

Decreasing the duration of the pulse increases its bandwidth. When this bandwidth encompasses the two exit channels, it induces a strong overlap effect that will strongly influence the photoelectron angular distribution. The importance of this effect depends on the relative values of the laser bandwidth and of the spectral separation between the two final ro-vibrational levels. As seen in Fig.\,\ref{fig:Photoelectron_Spectrum}, in the present scenario the peak corresponding to the channel $(v_+=0,N^+=0)$ is predominant because the rotational constants of the E-state of Li$_2$ and that of the ground  electronic state of Li$_2^+$ are very similar \cite{PRA.74.033407}. 

\begin{figure}[ht!]
\includegraphics[width=8.6cm]{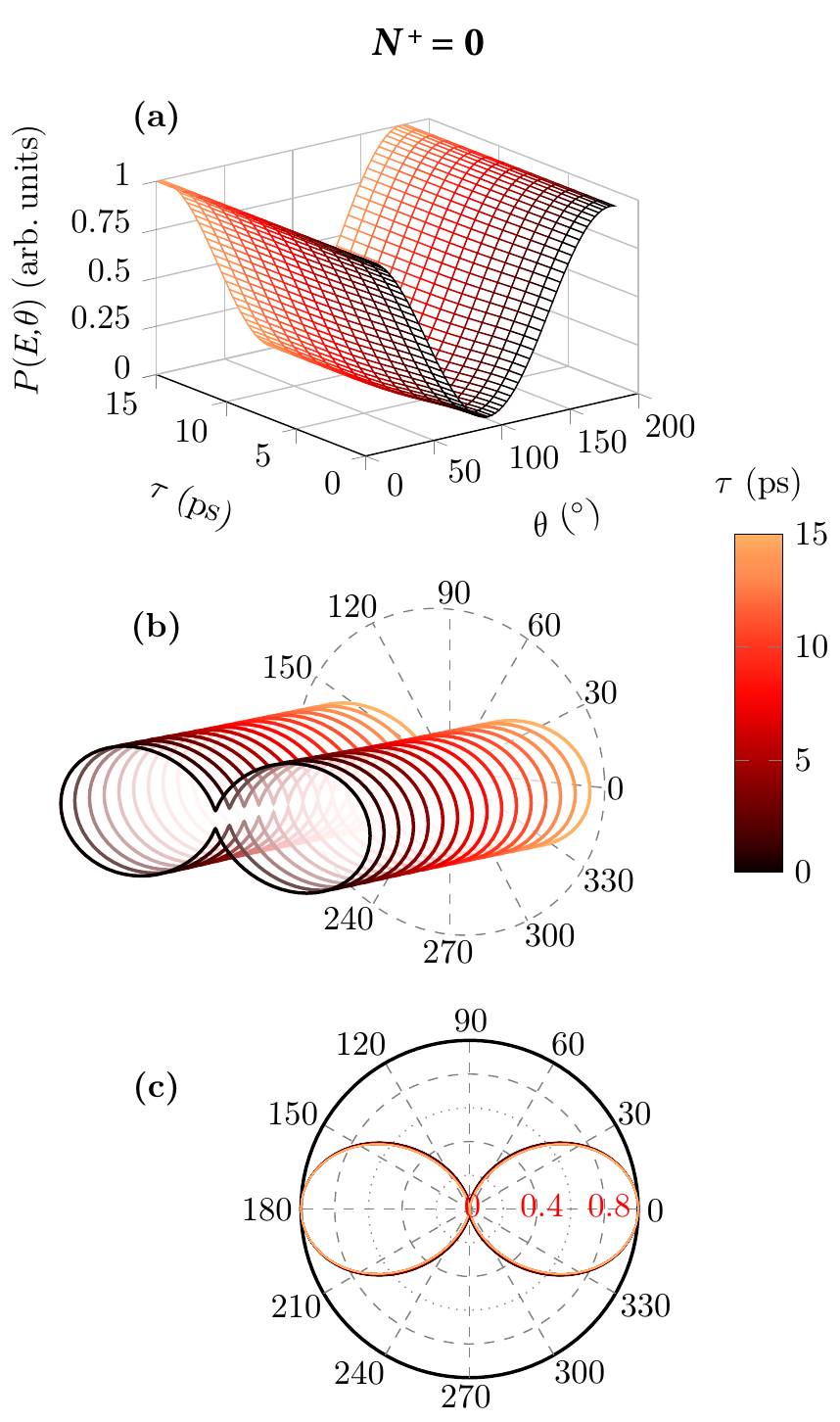}
\caption{(Color online) Photoelectron angular distribution spectra calculated using the initial $(v=0, N=0)$ rovibrational state for the main ion exit channel $(v_{+}=0, N^+=0$), which corresponds to the energy $E=4671.6$\,\wn, as a function of the ejection angle $\theta$ (varied from $0^{\circ}$ to $180^{\circ}$) and of the pulse duration $\tau$. Upper panel (a): Angular distribution spectrum $P(E,\theta)$ in a Cartesian view. Central panel (b): In a binocular view. Lower panel (c): In a polar view. The total pulse duration is varied from $\tau=50$\,fs to  $\tau=15$\,ps. The laser wavelength is $\lambda=532$\,nm (laser Nd:YAG).}
\label{fig:ANG_0_1}
\end{figure}

\begin{figure}[ht!]
\includegraphics[width=8.6cm]{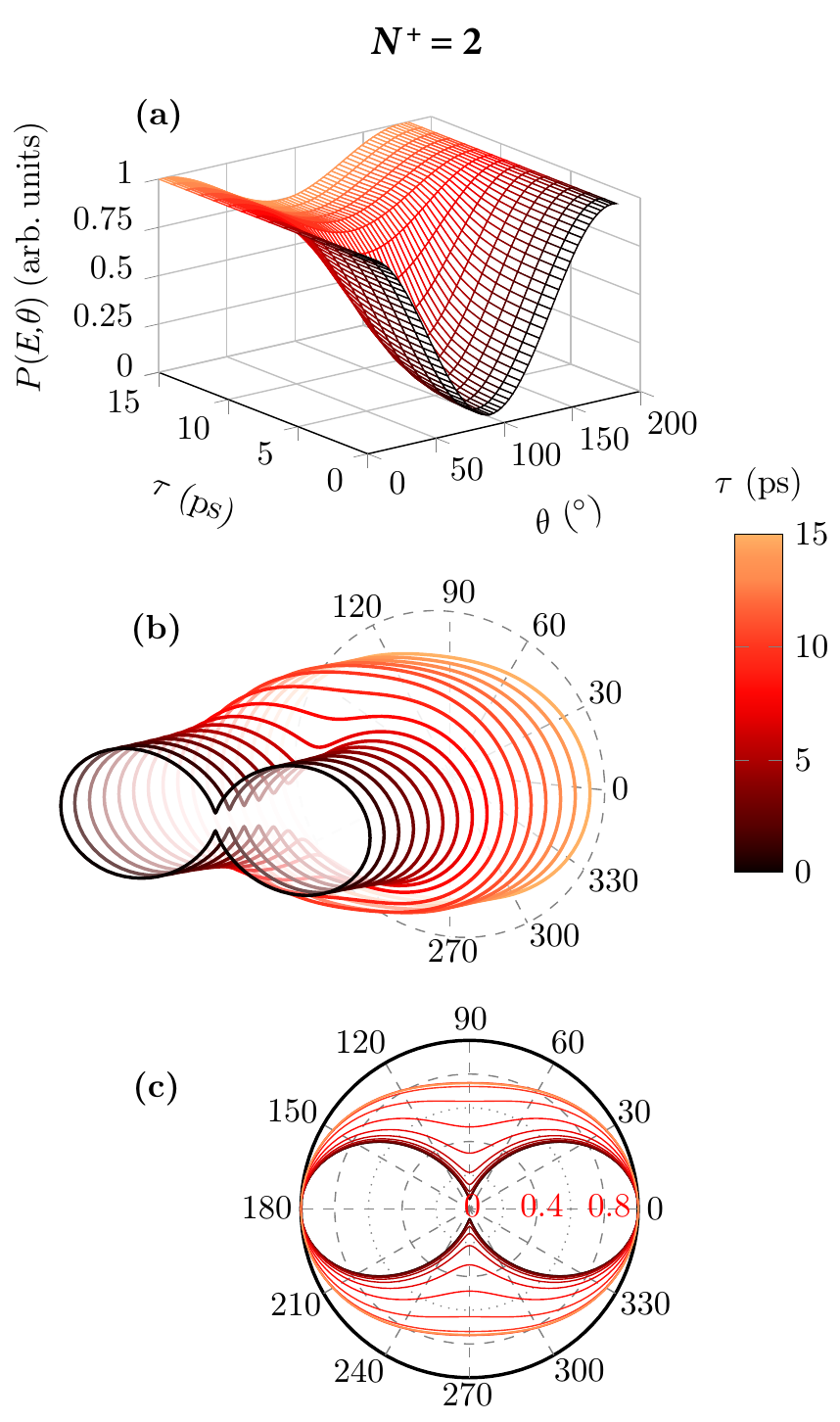}
\caption{(Color online) Photoelectron angular distribution spectra calculated using the initial $(v=0, N=0)$ rovibrational state for the secondary ion exit channel  $N^+=2$, which corresponds to the energy $E=4668.5$\,\wn, as a function of the ejection angle $\theta$ (varied from $0^{\circ}$ to $180^{\circ}$) and of the pulse duration $\tau$. Upper panel (a): Angular distribution spectrum $P(E,\theta)$ in a Cartesian view. Central panel (b): In a binocular view. Lower panel (c): In a polar view. The total pulse duration is varied from $\tau=50$\,fs to  $\tau=15$\,ps. The laser wavelength is $\lambda=532$\,nm (laser Nd:YAG).}
\label{fig:ANG_0_2}
\end{figure}

In Figs.\,\ref{fig:ANG_0_1} and \ref{fig:ANG_0_2}, we display the angular distribution spectra for the specific energies corresponding to the principal peak $(v_+=0, N^+=0)$ and to the secondary peak $(v_+=0, N^+=2)$, respectively. In each figure, the information are shown in three ways. In panel (a), the photoionization probability $P(E,\theta)$ as function of $\theta$, the angle between the polarization vector of the photon beam and the propagation vector of the ejected photoelectrons, for different values of the pulse duration. In panel (b), the  photoionization probability is plotted in polar coordinates for the same values of pulse durations as in panel (a), where short duration curves are plotted in front (black), while long duration curves are plotted in the back (yellow) of the 3D picture. The shape of this figure suggests to label this type of representation as a ``binocular'' view of the angular distribution. Finally, in panel (c), a front projection of panel (b) is performed.

The effect of the pulse duration on the angular distribution is strikingly different in both cases. For the main ($v_+=0, N^+=0$) exit channel ($E=4671.6$\,\wn, Fig.\,\ref{fig:ANG_0_1}), the shape of the distribution is independent of the pulse duration. Moreover, the distribution is peaked towards $\theta=0^{\circ}$ and $180^{\circ}$. The photoelectrons are thus predominantly ejected in the direction parallel to the photon polarization. However, in Fig.\,\ref{fig:ANG_0_2} corresponding to the minor $(v_+=0,N^+=2)$ exit channel ($E=4668.5$\,\wn), the photoelectron angular distribution evolves from a peaked anisotropic shape similar to the preceding case for short pulses $(\tau < 5\,\mathrm{ps})$, to a quasi-isotropic shape for longer pulses $(\tau > 10\,\mathrm{ps})$. This behavior results from the overlap of the two exit channels: when the pulse is short enough so that its bandwidth is larger than the separation between to two rovibrational levels, the first channel $(v_+=0,N^+=0)$ contributes to the signal and eventually dominates, even at the kinetic energy, which normally corresponds to the minor channel. 

To go forward in the investigation of the angular distribution at this point, we move to the analysis of the evolution of the asymmetry parameter $\beta$ with the pulse duration. We recall that $\beta$ is defined from the angular differential photoionization cross section as
\begin{equation}
\frac{d\sigma}{d\Omega}=\frac{\sigma}{4\pi}\;\Big[1+\beta\,P_{2}(\cos\theta)\Big]\,,
\label{eq: cross section}
\end{equation}
where $\sigma$ is the total photoionization cross section and $P_2(x)=(3x^2-1)/2$ is the Legendre polynomial of order 2. Using the notations adopted in the present study, we can write
\begin{equation}
P(E,\theta) \propto \Big[A+B\cos^2\theta\Big]\,,
\end{equation}
where $\beta$ can be deduced unambiguously from the relation
\begin{equation}
\beta = \frac{2B}{3A+B}\,.
\label{eq:beta2}
\end{equation}

We remind that  $\beta$, which can take values between $-1$ and 2, gives a global idea on the shape of the angular distribution of photoelectrons. For instance, $\beta=0$ corresponds to an isotropic distribution, while $\beta=2$ and $-1$ correspond to peaked anisotropic distributions along the light polarization axis and its perpendicular axis, respectively. It can be easily seen from Fig.\,\ref{fig:ANG_0_1}(b) that in this case $\beta$ is close to 2 and that in Fig.\,\ref{fig:ANG_0_2}(b) $\beta$ evolves from a value close to 2 for short pulses to a value close to 0 for long pulses. Indeed, the photoionization probabilities plotted in these figures are proportional to the differential photoionization cross section of Eq.(\ref{eq: cross section}). The aim of section\,\ref{sec:extr} is to try to understand this evolution of $\beta$ as a function of energy and pulse duration.

For a given energy  corresponding to one of the two exit channels and for a given pulse duration, we extract from Fig.\,\ref{fig:ANG_0_1}(a) or Fig.\,\ref{fig:ANG_0_2}(a), the extremal values of $P(E,\theta)$ as a function of $\theta$, namely $P_{max}=P(E,90^{\circ})$ and $P_{min}=P(E,0^{\circ})$. One can easily show that the ratio $\rho=P_{min}/P_{max}$ is directly related to $\beta$ through
\begin{equation}
\beta=\frac{\rho-1}{\rho/2+1}\,.
\end{equation}
This very simple formula allows an accurate estimation of the asymmetry parameter for each energy and pulse duration.

\begin{figure}[t!]
\includegraphics[width=8.6cm]{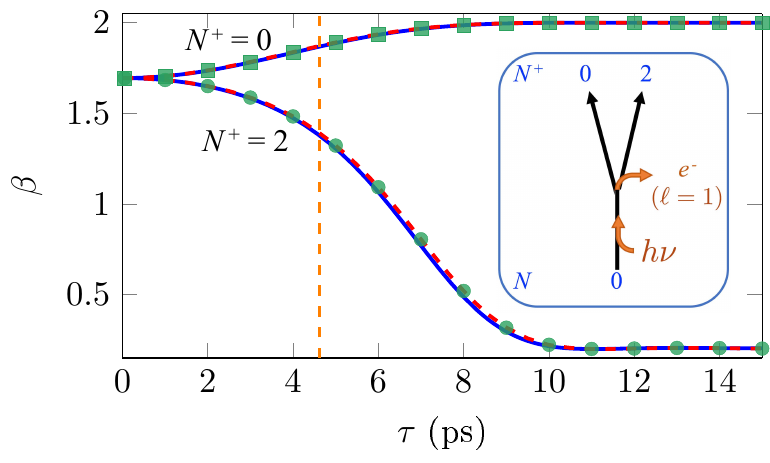}
\caption{Asymmetry parameter $\beta$ as a function of the pulse duration varying from $\tau=50$\,fs to  $\tau=15$\,ps for the two peaks corresponding to $N^+=0$ ($E=4671.6$\,\wn) and $N^+=2$ ($E=4668.5$\,\wn). It includes the  full quantum simulation (green symbols) and 2 curves: one for the analytical model with the exact parameters (solid blue lines), one for the numerical fit (dashed red lines). The inset shows the two accessible ionization pathways in competition. See text for details.}
\label{fig:BETA_0_l0}
\end{figure}

In Fig.\,\ref{fig:BETA_0_l0}, we display the asymmetry parameter $\beta$  as a function of the pulse duration for the two exit channels $N^+=0$ and $N^+=2$ corresponding to $E=4671.6$\,\wn $\,$ and  $E=4668.5$\,\wn, respectively. Three sets of data are plotted but we concentrate at the moment on the symbols (green squares and circles), which are the values extracted from our time-dependent quantum calculation. The other estimations correspond to an analytical model (solid blue lines) and to a numerical fit (dashed red lines) that we will discuss in section\,\ref{sec:extr}. We can already note that the agreement between the three sets of results is very good, demonstrating that the analytical model and the numerical fit allow for a good description of the dynamics of the system. For very short pulses, the two curves (for both exit channels) converge to the same limit $\beta \simeq 1.69$. This behavior is correlated with the overlap effect that we have already discussed. Finally, as the pulse duration increases, the two curves separate and reach their asymptotic values $\beta= 0.2$ and $\beta=2$ for the $N^+=2$ and $N^+=0$ channels, respectively. These exact values can be derived easily. Indeed, for a given $N^+$ the photoionization probability as a function of $\theta$, deduced from Eq.\,(\ref{eq:PNplus(E,theta)simple}), can be written in the CW limit as
\begin{equation}
P_{N^+}(E,\theta) \propto
\sum_{m} \,|Y_{\ell m}(\theta,\phi)|^2\,|{\cal{M}}_{\ell m}^{N,N^+}|^2\,,
\end{equation}
where ${\cal{M}}_{\ell m}^{N,N^+}$ is the electric dipole matrix element between the initial and final states \cite{PRA.74.033407}. For each exit channel, the value of $\beta$ can be deduced unambiguously from Eq.\,(\ref{eq:beta2}). Indeed, for $N^+=0$ we find $P_{0}(E,\theta) \propto \cos^2\theta$, giving $A=0$ and therefore $\beta=2$. On the other hand, for $N^+=2$, we have $P_{2}(E,\theta) \propto (3 + \cos^{2}\theta)$, {\it i.e.} $A=3$, $B=1$ and thus $\beta=1/5=0.2$, as expected.

\begin{figure}[t!]
\includegraphics[width=8.6cm]{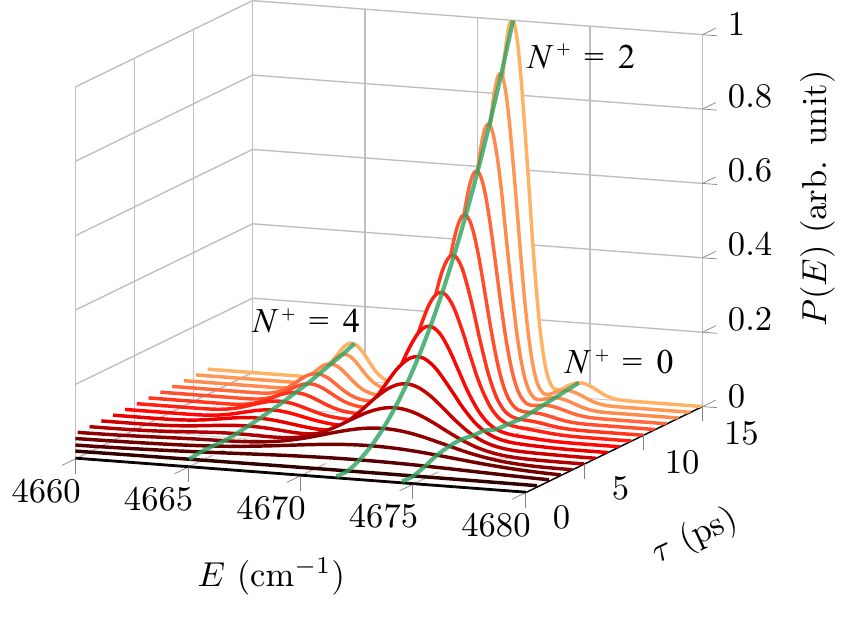}
\caption{(Color online) Photoelectron spectra calculated with the initial rovibrational level $(v=0,N=2)$ as a function of the electron kinetic energy $E$ and of the pulse duration $\tau$. The laser wavelength is $\lambda=532$\,nm (laser Nd:YAG).}
\label{fig:Photoelectron_Spectrum2}
\end{figure}

\subsubsection{Excited rotational case $N=2$}

The main difference when starting initially from the $(v=0,N=2)$ excited rovibrational level is that there are three exit channels corresponding to the three ion rotational quantum numbers $N^+=0$, 2 and 4 when a $p$ electron is emitted. The overlap effect discussed in the previous Section is therefore expected to be more rich and complex, especially for short pulse durations.

\begin{figure}[t!]
\includegraphics[width=8.6cm]{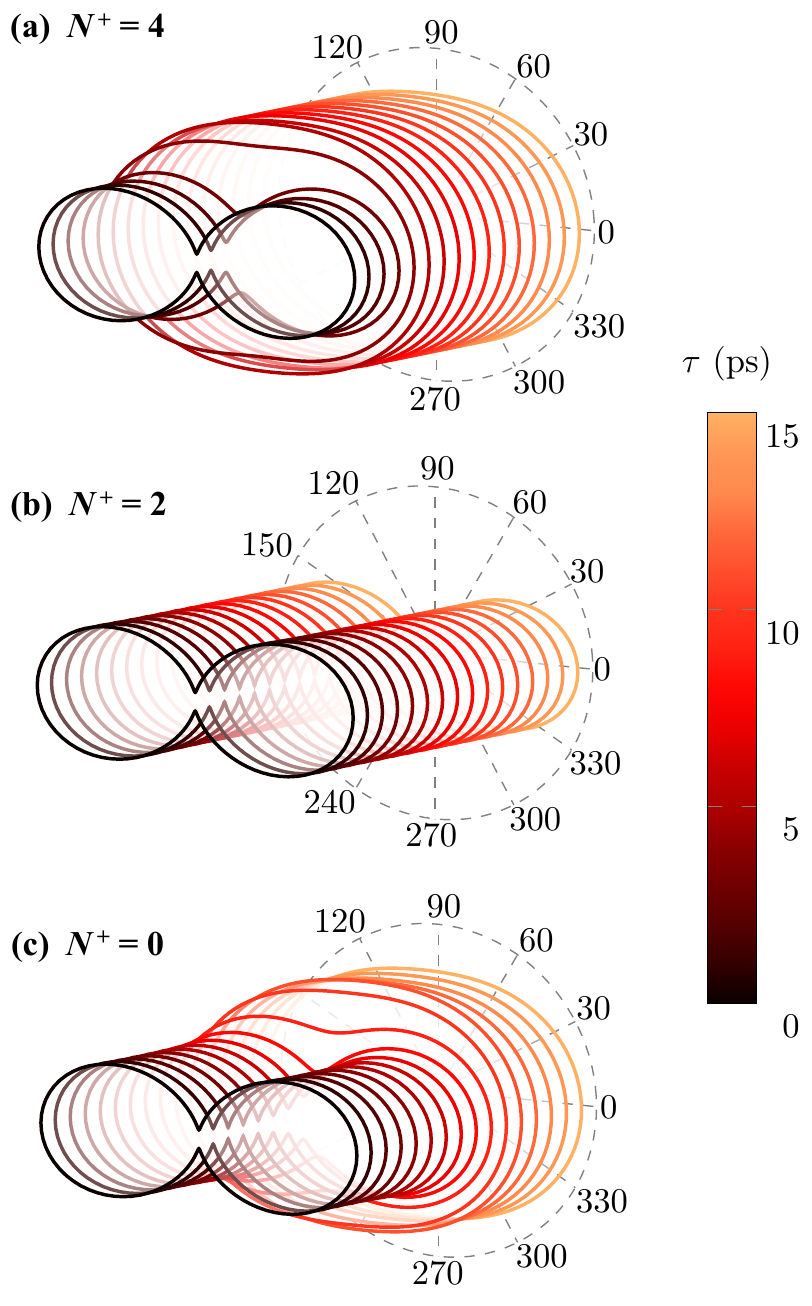}
\caption{(Color online) Binocular view of the angular distribution spectra for the initial state $(v=0,N=2)$ as a function of $\theta$ and a total pulse duration $\tau$. Upper panel (a):  Exit channel $N^+=4$ corresponding to the energy $E=4664$\,\wn. Central panel (b): Exit channel $N^+=2$ corresponding to the energy $E=4671$\,\wn. Lower panel (c): Exit channel $N^+=0$ corresponding to the energy $E=4674$\,\wn. The total pulse duration is varied from $\tau=50$\,fs to $\tau=15$\,ps. The laser wavelength is $\lambda=532$\,nm (laser Nd: YAG).}
\label{fig:ANG_2}
\end{figure}

Fig.\,\ref{fig:Photoelectron_Spectrum2} shows the photoionization probability as a function of the electron kinetic energy for various pulse durations. In a way similar to Fig.\,\ref{fig:Photoelectron_Spectrum}, for very short pulses, the spectra show a broad structureless single  peak. However, its compound structure splits into three well-separated peaks  for long pulse durations. Moreover, the principal peak at $E= 4671$\,\wn  $\,$ corresponds to the main exit channel $N^+=2$, in accordance with the propensity rule $N^+=N$. The two other exit channels give rise to two satellite peaks on both sides of the main peak, namely a lower energy peak around $E=4664$\,\wn  $\,$ corresponding to $N^+=4$ and a higher energy peak near $E=4674$\,\wn  $\,$ for $N^+=0$.

In Fig.\,\ref{fig:ANG_2}, the photoelectron angular distributions at these three energies are plotted in the previously defined binocular view as a function of the ionization angle $\theta$ for various pulse durations. The three channels $N^+=4$, $N^+=2$ and $N^+=0$ are shown in panels (a), (b) and (c), respectively. Here again, the predominant channel $N^+=2$, is almost not affected by the pulse duration whereas the satellite channels present angular distributions that evolve from peaked shapes at short durations to quasi-isotropic shapes for very long pulses. The evolution shown in Fig.\,\ref{fig:ANG_2}(c) for $N^+=0$ is progressive and similar to that of Fig.\,\ref{fig:ANG_0_2}(b). However, in the case of the $N^+=4$ exit channel, the transition from an anisotropic photoelectron angular distribution to a quasi-isotropic distribution is sharp, occurring around $\tau\simeq 3$\,ps.

The asymmetry parameter $\beta$ for the three channels is displayed in Fig.\,\ref{fig:BETA_0_l2} as a function of the pulse duration $\tau$. It shows an evolution similar to that of Fig.\,\ref{fig:BETA_0_l0}, which was showing the evolution of the same parameter for the initial rovibrational state $N=0$. For very short pulses, the three curves become degenerate at the value $\beta \simeq 1.69$, while they fully separate in the intermediate range 1\,ps\;$\leqslant\tau\leqslant$\;10\,ps, before reaching their asymptotic values $\beta \simeq 1.9$ for the principal exit channel $N^+=2$ ($E=4671$\,\wn) and $\beta \simeq 0.2$ for secondary exit channels $N^+=0$ ($E=4674$\,\wn) and $N^+=4$ ($E=4664$\,\wn). Here again, we have a very good agreement between the values calculated using the time-dependent quantum simulation, the analytical model and the numerical fit for all exit channels. This analytical model and the numerical fit will be described in the next section.

\subsection{Extraction of spectroscopic molecular parameters}
\label{sec:extr}

\begin{figure}[t!]
\includegraphics[width=8.6cm]{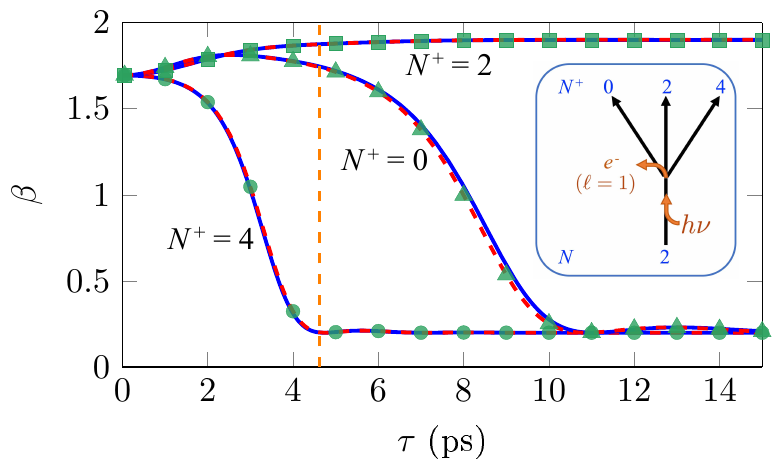}
\caption{Asymmetry parameter $\beta$ as a function of the pulse duration from $\tau=50$\,fs to $\tau=15$\,ps for the three peaks corresponding to $N^+=0$ ($E=4674$\,\wn), $N^+=2$ ($E=4671$\,\wn) and $N^+=4$ ($E=4664$\,\wn). It includes the  quantum simulation with green symbols and 2 curves: one for the analytical model with the exact parameters (solid blue lines) and one for the numerical fit (red dashed lines). The inset shows the three accessible ionization pathways in competition. See text for details}
\label{fig:BETA_0_l2}
\end{figure}

In contrast with the CW case, for short pulse durations, a given electron kinetic energy $E$ may be associated with different exit channels $N^+$. This effect is simply due to the increased bandwidth of the pulse which leads to non-zero contributions of different $N^+$ channels, as seen in Eq.\,(\ref{eq:P(E,theta)simple}). The strong variation of the asymmetry parameter $\beta$ with the pulse duration can thus be easily rationalized from the evidence that the electron angular distribution can be written as an incoherent sum over the different exit channels as written in Eq.\,(\ref{eq:P(E,theta)simple}). As a consequence, the differential ionization probability can be written as
\begin{equation}
P(E,\theta) \propto \Big[1+\beta(\tau)\,P_{2}(cos\theta)\Big]
\end{equation}
for an arbitrary pulse duration $\tau$, where $\beta(\tau)$ is given by a weighted average of the CW $\beta$-values of the different individual exit channels $N^+$
\begin{equation}
\beta(\tau) = \frac{\sum_j P_j\,\tilde{f}_j\,\beta_j}{\sum_j P_j\,\tilde{f}_j}\,,
\label{eq:beta_model}
\end{equation}
where $P_j$ is the total ionization probability in the exit channel number $j$, $\tilde{f}_j$ denotes the field amplitude at the frequency corresponding to the energy of the exit channel number $j$, and $\beta_j$ is the asymptotic CW value of the asymmetry parameter of channel $j$. Note that since the probabilities $P_j$ are proportional to the square of the electric dipole moment $\mu_j^2$ between the initial state and a given final state $j$, Eq.\,(\ref{eq:beta_model}) can equivalently be written as
\begin{equation}
\beta(\tau) = \frac{\sum_j \mu_j^2\,\tilde{f}_j\,\beta_j}{\sum_j \mu_j^2\,\tilde{f}_j}\,.
\label{eq:beta_model2}
\end{equation}
We have been using Eq.\,(\ref{eq:beta_model2}) to compare, in Figs.\,\ref{fig:BETA_0_l0} and \ref{fig:BETA_0_l2}, the result obtained from the numerical solution of the time-dependent Schr\"odinger equation (solid green squares and circles) with the analytical model (solid blue lines) and the agreement is extremely good.

Another advantage of the knowledge of Eq.\,(\ref{eq:beta_model2}) is that it allows, from a limited numbers of measurements performed with ultrashort laser pulses, to extract the asymptotic CW asymmetry parameters $\beta_j$ of the different exit channels as well as the relative electric dipole moments $\mu_j^2$ by using a simple numerical fitting procedure. We have shown the results of such a numerical fit using Eq.\,(\ref{eq:beta_model2}) in Figs.\,\ref{fig:BETA_0_l0} and \ref{fig:BETA_0_l2} in red dashed lines. The fitting parameters are the values of $\beta_j$ and $\mu_j^2$ since we consider that for a given pulse duration $\tau$ the Fourier transform $\tilde{f}_j$ of the electric field $E(t)$ at the frequency corresponding to each exit channel is known. The agreement of the numerical fit with the numerical solution of the time-dependent Schr\"odinger equation is excellent. For example, using the 5 values of $\beta$ shown with green symbols in Fig.\,\ref{fig:BETA_0_l0} for $\tau \leqslant 4.5$\,ps (on the left of the dashed vertical line), we obtained $\beta_0=2.0000$ instead of 2 and $\beta_2=0.19995$ instead of 0.2 for the initial state $N=0$. One can note in Fig.\,\ref{fig:Photoelectron_Spectrum} that for these very short pulses the rotation is not yet resolved in the kinetic energy spectrum, which shows a single broad peak. Indeed, for such ultrashort pulses, the pulse duration is smaller than the rotational period of both \lit and Li$_{2}^{+}$. Similarly, using the 5 values of $\beta$ shown with green symbols in Fig.\,\ref{fig:BETA_0_l2} for $\tau \leqslant 4.5$\,ps, we obtained $\beta_0=0.2099$ instead of 0.2 (error $<$ 5\%), $\beta_2=1.897$ instead of 1.9 (error $<$ 1\%) and $\beta_4=0.2075$ (error $<$ 4\%) instead of 0.2 for the initial state $N=2$. The numerical values obtained with this fitting procedure for the relative electric dipole moments $\mu_j^2$ are also correct within 5\%.

With a limited number of measurements performed with very short laser pulses it is thus possible to extract accurate values of some spectroscopic molecular parameters such as the asymmetry parameters of the different exit channels, thanks to the simplicity of Eq. (\ref{eq:beta_model2}), and therefore to the efficiency of the associated numerical fit.

\section{Conclusion}
\label{sec:conclusion}

In this paper, we have presented a theoretical study of the short pulse ionization of the \lit molecule in the fs to ps regime, solving the time-dependent Schr\"odinger equation for both the nuclear and the electronic dynamics. We have concentrated our study on the calculation of the energy-resolved angular distributions of the emitted photoelectrons. We have shown that when the ionizing pulse is much shorter than the typical rotational period of the molecule the kinetic energy distribution is characterized by a single broad peak which comprises the contribution of the different ion rotational levels involved. In this ultrafast regime, one would normally expect that it is impossible to extract spectroscopic molecular parameters such as the asymmetry parameters $\beta$ associated with each individual ion rotational level. In this paper we have shown however that it is in fact possible to extract such values from a limited number of measurements performed with different pulse durations. These values can be obtained from a numerical fitting procedure and we have shown that the accuracy of this procedure is of the order or better than 5\%. In the future, we are planning to extend the present model to the study of the dissociative ionization of Li$_{2}$. This would allow to extract more information from coincidence measurements \cite{Dowek,MP.110.131}.

\section*{Acknowledgments}

We acknowledge the use of the computing cluster GMPCS of the LUMAT federation (FR 2764 CNRS). We acknowledge CINES, France for providing access and support to their computing platform Occigen under project AP-010810188. MT acknowledges an invited Professor position at the Paris-Sud University, during which part of this work has been conducted. 

%

\end{document}